\documentclass[3p,times,procedia]{elsarticle}
\usepackage{nupha_ecrc}

\volume{00}

\firstpage{1}

\journalname{Nuclear Physics A}

\runauth{J.-F. Paquet for the JETSCAPE Collaboration}

\jid{nupha}

\jnltitlelogo{Nuclear Physics A}

\usepackage{amssymb}

\usepackage[figuresright]{rotating}

\begin{document}

\begin{frontmatter}



\dochead{XXVIIIth International Conference on Ultrarelativistic Nucleus-Nucleus Collisions\\ (Quark Matter 2019)}

\title{Revisiting Bayesian constraints on the transport coefficients of QCD}


\author{Jean-Fran\c{c}ois Paquet for the JETSCAPE Collaboration}

\address{Department of Physics, Duke University, Durham, NC 27708, USA}

\begin{abstract}
Multistage models based on relativistic viscous hydrodynamics have proven successful in describing  hadron measurements from relativistic nuclear collisions.
These measurements are sensitive to the shear and the bulk viscosities of QCD and provide a unique opportunity to constrain these transport coefficients.
Bayesian analyses can be used to obtain systematic constraints on the viscosities of QCD, through methodical model-to-data comparisons.
In this manuscript, we discuss recent developments in Bayesian analyses of heavy ion collision data.
We highlight the essential role of closure tests in validating a Bayesian analysis before comparison with measurements. 
We discuss the role of the emulator that is used as proxy for the multistage theoretical model.
We use an ongoing Bayesian analysis of soft hadron measurements by the JETSCAPE Collaboration as context for the discussion.
\end{abstract}


\end{frontmatter}


\section{Multistage models of heavy ion collisions}
\label{sec:model}

The quark-gluon plasma produced in heavy ion collisions expands, cools down and reconfines into hadrons within $\sim10$~fm/c. This rapid spacetime evolution of the plasma can be described with relativistic viscous hydrodynamics, starting from a time $\tau \sim$ 0.1--1~fm/c.
The phase that precedes this hydrodynamic regime ($\tau\lesssim$0.1--1~fm/c) is the ``pre-equilibrium phase''; the appropriate theoretical description of this early stage of the plasma is still under active investigation~\cite{Gale:2013da:plus}. 
The stage that follows the hydrodynamic regime ($\tau \gtrsim 10$~fm/c) is described microscopically with a transport model of hadrons, after which hadronic observables are evaluated and compared to measurements~\cite{Gale:2013da:plus}.
Because of the complex nature of the multistage model, comparisons with data 
are used to study the shear and bulk viscosities of QCD
at the same time as constraining other model parameters
(related for example to the ``pre-equilibrium phase'').
The number of unknowns involved generally makes it necessary to use a combination of different measurements to obtain usable constraints~\cite{Bernhard:2019bmu,Sangaline:2015isa,Novak:2013bqa}. 
Bayesian analyses provide a methodical way of constraining the model parameters from such an array of measurements.


In what follows, we discuss the basis of the Bayesian analysis framework used by the JETSCAPE Collaboration for its analysis of soft hadrons~\cite{jetscape_soft_probes_bayes}.
Two primary aims of this analysis are (i) constraining the properties of the quark-gluon plasma with soft hadron measurements from multiple collision systems, and (ii) provide the best possible description of the quark-gluon plasma such that quantitative studies of hard observables can be performed. The present work focuses on the first objective.


\section{Bayesian analyses}

The power of Bayesian analyses is to make model-to-data comparisons systematic while making efficient use of both the measurements and their uncertainties. To do so, we first promote measurements and model calculations to the status of probability distributions (Gaussian in our case). The central value of the measurement/calculation is the mean of the probability distribution, while the width of the distribution is related to their uncertainties. 
These uncertainties can be statistical, systematic or numerical.
The result of the Bayesian analysis can be seen as combining the probability distributions of the model and the data, obtaining probabilistic constraints on the model parameters~\cite{tarantola2005inverse}: the likelihood 
that a set of parameters 
is consistent with the measurements. 


Bayesian analyses are flexible, as they can combine the constraining power of a large number of measurements onto a significant number of model parameters. 
On the other hand, Bayesian analyses are numerically expensive, as they require knowledge of the model's prediction across a wide range of model parameters.

The number of model parameters can be large.
The shear and bulk viscosities of QCD themselves, parametrized through their ratio to entropy density (respectively $\eta/s$ and $\zeta/s$), are non-trivial functions of temperature, and possibly chemical potentials. Even with strong assumptions about their functional form, multiple parameters can be necessary for each viscosity.
There are also other model parameters associated with unknowns in the multistage model of heavy ion collisions.
The JETSCAPE analysis~\cite{jetscape_soft_probes_bayes} contains 17 parameters, including (i) 7 parameters for the pre-equilibrium stage of the collision, and (ii) 4 parameters each for $(\eta/s)(T)$ and $(\zeta/s)(T)$.

Ultimately, the Bayesian analysis quantifies to what extent different combinations of model parameters are consistent with the measurements and the theoretical and experimental uncertainties. Doing so requires evaluating the theoretical model across the parameter space. This exploration of the parameter space is performed with a Markov chain Monte Carlo approach. For our 17-dimensional parameter space~\cite{jetscape_soft_probes_bayes},
we use $\mathcal{O}(10^6)$ Markov steps to probe the likelihood.
Each step in theory requires a complete minimum bias simulation of the multistage model, which can represent $\mathcal{O}(10^3$--$10^4)$ executions of the multistage model.
To mitigate the numerical cost of the Bayesian analysis, we perform the Markov chain Monte Carlo on proxies for the theoretical model: ``emulators''. 

\section{Emulators}

The ultimate output of the model is a set of observables (multiplicity, momentum anisotropy, \ldots) which can eventually be compared with experimental data.
Emulators are direct maps between the model parameters and these calculated observables.
This mapping is constructed by evaluating the model for a small sample of parameters across the parameter space. The emulators interpolate the observables from these sampled parameter points to predict the value of the observables across the parameter space. There are thus multiple emulators corresponding to the multiple observables of interest.\footnote{In practice, we do not use a separate emulator for each observable. Instead, observables are grouped together according to their sensitivity to the model parameters, using principal component analysis, and emulators are used to emulate a subset of dominant principal components. Details can be found in Ref.~\cite{Bernhard:2018hnz}.}
Because emulators are never an exact representation of the model, they introduce an additional source of uncertainty in the Bayesian analysis.

With our 17-dimensional parameter space~\cite{jetscape_soft_probes_bayes}, we use $\mathcal{O}(10^3)$ samples of the parameter space to train the emulators. The sampling is performed with the Latin hypercube algorithm to distribute the limited number of samples across the parameter space.
For the emulators themselves we use ``Gaussian processes'' (see Ref.~\cite{Bernhard:2018hnz} and references therein for details). Just like measurements and the theoretical model's predictions are considered probability distributions, emulators are high-dimensional probability distributions.
The $\mathcal{O}(10^3)$  samples of the parameter space are ``anchor points'' in the parameter space where the emulators' probability distributions are constrained by the model's prediction.\footnote{In our current analysis, the emulators are in fact not exactly the probability distribution of the model at the ``anchor points''. This is an implementation choice in which the statistical uncertainty of the model is estimated by adding a noise term to the emulators. See Ref.~\cite{Bernhard:2018hnz} for details.} Away from these ``anchor points'', the emulator's probability distribution is an interpolation. As such, the ``width'' of this probability distribution increases away from the parameter ``anchor points'', to reflect the interpolation uncertainty of the emulator.
We generally find a $5-10\%$ emulator uncertainty when training the emulator with $\mathcal{O}(10^3)$ parameter samples~\cite{jetscape_soft_probes_bayes}.


Validating the emulator is essential, as any defect in the emulator would be interpreted as a physical feature of the model.
We explored using two methods: (i)  compare the emulator's predictions to model calculations that have not been used to train the emulator, and (ii) train the emulator on all model calculations except one and comparing its prediction to this excluded model calculation.
However, here, we favor an additional form of validation that simultaneously validates the entire Bayesian analysis: closure tests.

\section{Closure tests}
\label{sec:closure}

A closure test is simply a Bayesian analysis performed on model calculations. First, hadronic observables $\mathcal{R}$ are calculated with a chosen set of model parameters $\mathbf{p}$. Separately, an emulator is trained on a sample of model calculations,  as described in the previous section. This emulator is then used to perform a Bayesian analysis on the hadronic observables $\mathcal{R}$. If the emulator were a perfect representation of the model, and if there were no uncertainties at all in the problem, the Bayesian analysis should generally
recover the model parameters $\mathbf{p}$: the likelihood would be sharply peaked around $\mathbf{p}$. 

\begin{figure}
	\centering
	\includegraphics[width=0.35\linewidth]{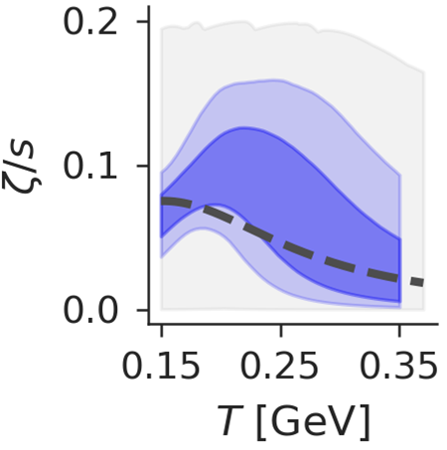} \llap{
		\parbox[b]{2.1in}{(a)\\\rule{0ex}{0.05in}
}}
	\hspace{0.3cm}
	\includegraphics[width=0.35\linewidth]{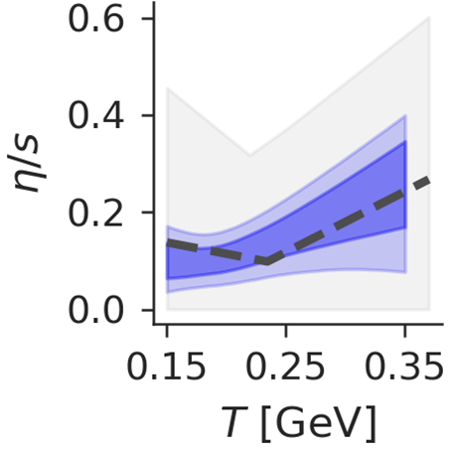} \llap{
		\parbox[b]{2.1in}{(b)\\\rule{0ex}{0.05in}
}}
	\caption{(a) Closure test on the temperature dependence of (a) $(\zeta/s)(T)$ and (b) $(\eta/s)(T)$, respectively the bulk and shear viscosity to entropy density ratio. The dashed line is the actual value of $(\zeta/s)(T)$ and $(\eta/s)(T)$ while the light and dark blue shaded areas are the $60$\% and $90$\% confidence intervals of the Bayesian analysis. The gray area is an estimate of the prior. See text for additional details.}
	\label{fig:closure}
\end{figure}

Because every calculation has numerical and statistical uncertainties, and because emulators also have uncertainties, closure tests are non-trivial. 
An example of a closure test is shown in Fig.~\ref{fig:closure} for the temperature dependence of $(\zeta/s)(T)$ and $(\eta/s)(T)$, the bulk and shear viscosity to entropy density ratio.
The dashed black line is a certain choice of $(\zeta/s)(T)$ and $(\eta/s)(T)$ (parameter ``$\mathbf{p}$'') from which hadronic observables\footnote{This example used the following observables for Pb-Pb collisions at $\sqrt{s_{NN}}=2.76$~TeV: transverse energy, charged hadron $dN/d\eta$, pion, kaon and proton $dN/dy$ and $\langle p_T \rangle$, charged hadron $v_{2/3/4}\{2\}$ and charged hadron mean $p_T$ fluctuations.} $\mathcal{R}$ were calculated. The light and dark blue shaded areas are the $60$\% and $90$\% confidence intervals of the Bayesian analysis performed on the hadronic observables $\mathcal{R}$.

The first conclusion from this closure test is that both the correct $(\zeta/s)(T)$ and  $(\eta/s)(T)$ are within the high-probability confidence intervals identified by the Bayesian analysis, as they should be. In this sense, this closure test is successful. Nevertheless, the confidence interval is generally broad. This indicates that $(\zeta/s)(T)$ and $(\eta/s)(T)$ are difficult to constrain. To narrow the confidence interval, one may need a combination of smaller emulator uncertainties, additional observables and reduced uncertainties on the observables.

The width of the confidence interval in the closure test is important: 
if the Bayesian analysis were performed on experimental data using the same emulator, the constraints on $(\zeta/s)(T)$ and $(\eta/s)(T)$ are expected to be similar in quality to what is seen in Fig.~\ref{fig:closure}.
This is because experimental uncertainties will add to the emulator uncertainties, making it even more challenging to constrain the viscosities of QCD.
As such, before performing any comparisons with measurements, one can 
foresee
the extent to which a Bayesian analysis will be able to constrain a given set of model parameters.

Since closure tests validate at the same time all the other ingredients of the Bayesian analysis, we put forward that every Bayesian analysis should perform closure tests first before comparing with data. Systematic studies of each observable's dependence on the model parameters~\cite{Sangaline:2015isa} can also be performed using closure tests, to help clarify which observables hold the most potential to constrain the properties of QCD.
As shown by Fig.~\ref{fig:closure}, closure tests can also help determine what are the current limitations of the Bayesian analysis: improving the Bayesian analysis itself (for example, reducing emulator uncertainties) can be more important than adding observables or reducing experimental uncertainties.


\section{Outlook}
\label{sec:outlook}

Bayesian analyses are powerful tools for model-to-data comparisons. 
They can combine experimental constraints from a large number of observables in a systematic, reproducible fashion, yielding meaningful probabilistic constraints on model parameters.
Correlations in observables and in their uncertainties can be accounted to make efficient use of measurements~\cite{jetscape_hard_probes_bayes}.

Pioneering works have shown that Bayesian analyses can be used successfully  in heavy ion physics~\cite{Bernhard:2019bmu,Sangaline:2015isa,Novak:2013bqa}.
In this work, we discussed different ways to build upon these previous studies.
The role of the emulator in Bayesian analyses deserves scrutiny. Closure tests can play a major role in validating  Bayesian analyses, putting them on a surer footing before proceeding to comparison with experiments.
Other aspects of the analyses not discussed here also need to be reevaluated; this includes the role of parameter priors, as well as the important question of systematic theoretical uncertainties.

These aspects of Bayesian analysis of heavy ion measurements are currently being revisited in the ongoing soft matter calibration~\cite{jetscape_soft_probes_bayes} of the JETSCAPE Collaboration~\cite{Putschke:2019yrg}.
We are confident that the field as a whole will benefit from our methodical approach to Bayesian analyses, and from the lessons being drawn from our exploration of the topic.



\paragraph{Acknowledgments} 
This work was supported by the U.S.
Department of Energy (DOE) under
award number DE-FG02-05ER41367
and by the National Science
Foundation (NSF) under award
number ACI-1550300.

\bibliographystyle{elsarticle-num}
\bibliography{biblio}







\end{document}